\documentclass[a4paper]{ESASPCS13Style}
\usepackage{epsfig}

\begin{document}

\title{Spatially Resolved {\it STIS} Spectra of Betelgeuse's Upper Chromosphere and Circumstellar Dust Envelope}

\author{A. Lobel} \institute{Harvard-Smithsonian Center for Astrophysics, 60
  Garden Street, Cambridge, 02138 MA, USA}
\maketitle 

\begin{abstract}
The Hubble Space Telescope observed red supergiant 
Betelgeuse ($\alpha$ Ori) with the Space Telescope Imaging Spectrograph
to investigate the outer atmosphere from spatially resolved spectra.
We present a new set of seven high-resolution near-UV 
spectra observed with HST-$STIS$ in fall 2002 and spring 2003,
by scanning at chromospheric intensity peak-up position and 
six off-limb target positions up to three arcseconds away from the star. 
A small aperture is used to study and determine the thermal conditions and flow 
dynamics in the upper chromosphere and inner circumstellar dust envelope
of this important nearby cool supergiant (M2 Iab). 

We provide the first evidence for the presence of warm 
chromospheric plasma at least 3\arcsec\, away from Betelgeuse at 
$\sim$120 $\rm R_{*}$ (1 $\rm R_{*}$$\simeq$700 $\rm R_{\odot}$)
based on detailed spectroscopic observations of the Mg~{\sc ii} $h$ \& $k$ 
emission lines. Many other weak chromospheric emission lines as
Fe~{\sc ii} $\lambda$2716, C~{\sc ii} $\lambda$2327, Al~{\sc ii} ]
$\lambda$2669, and Fe~{\sc i} $\lambda$2823, are detected out to 
at least 1\arcsec\, in the spatially resolved $STIS$ observations. 
The recent spectra reveal that $\alpha$ Ori's upper chromosphere extends 
far beyond the circumstellar H$\alpha$ envelope of $\sim$5 $\rm R_{*}$, 
determined from previous ground-based images. 
The changes of shape of the detailed 
Mg~{\sc ii} line profiles observed in Betelgeuse's outer atmosphere are compared
with detailed Mg~{\sc ii} line profiles previously observed above the limb of
the average quiet Sun. The profiles of the Mg~{\sc ii} $h$ \& $k$, and the 
Si~{\sc i} resonance emission lines reveal a strong increase of asymmetry by 
scanning off-limb, signaling the outward acceleration of wind expansion in 
Betelgeuse's upper chromosphere beyond 200 mas ($\sim$8 $\rm R_{*}$). 

We discuss detailed radiative transfer models that fit the $STIS$ observations 
showing that the local kinetic gas temperature in the upper chromosphere
exceeds 2600~K. Our radiation transport models for the IR silicate dust emission
at 9.8 $\mu$m in the upper chromosphere show however that the ambient gas
temperature remains below 600 K to sustain the presence of dust grains.
Hence, the $STIS$ spectra of Betelgeuse's upper
chromosphere directly demonstrate that warm chromospheric plasma must co-exist 
with cool dusty plasma in its outer atmosphere.  

\end{abstract}

\section{Introduction}
Ongoing research of Betelgeuse's (\object{HD~39801}) extended chromosphere and dust envelope
has recently provided important advances in understanding the dynamics
and thermal structure of the atmospheres and winds of evolved stars. 
Detailed investigations of cool star atmospheres are much more 
intricate compared to hot stars because of enhanced atomic and molecular 
opacities and complicated dynamic activity (pulsation and convection), 
although late-type stars are one of the most important laboratories 
for stellar atmospheric physics. High-resolution spectroscopy 
of luminous cool stars provides fundamental information about the complex 
physical mechanisms that cause their exceptionally large mass-loss rates 
($\dot{M}$$\geq$$10^{-6}$$\,\rm M_{\odot}\,yr^{-1}$) by 
which these massive stars continuously replenish the interstellar medium with 
material that has been processed through nuclear fusion reactions, and from 
which many structures in circumstellar and interstellar environments
originate. Unlike the supersonic accelerating winds of hot supergiants, the mass-loss
mechanisms that drive the slower winds of cool supergiants like \object{Betelgeuse} are presently 
not well understood due to a persistent lack of high-quality ultraviolet 
spectra that probe the important physical properties of atmospheric regions 
located between the photosphere and the extended circumstellar dust envelope. 

Spatially resolved raster scans observed with $STIS$ using a small aperture across 
Betelgeuse's UV disk in HST Cycles 7 \& 8 reveal
subsonic oscillations of the inner chromosphere from radiative transfer fits 
to the detailed asymmetric profiles of the Si~{\sc i} $\lambda$2516 
resonance emission line (Lobel 2001, Lobel \& Dupree 2001). The near-UV spectra 
show long-term changes of gas movements in the lower chromosphere 
with pulsations of the deeper -presently unresolved- stellar photosphere. 
For the first time, these two-dimensional STIS spectra 
provided direct evidence that $\alpha$ Ori's inner chromosphere oscillates
non-radially, thereby occasionally exhibiting simultaneous up- and
down-flows. 

\section{HST-STIS Observations of 2002-2003}
Recent $STIS$ spectra of Betelgeuse have been observed in the fall of 2002 and spring
of 2003 for GO 9369 in HST Cycle 11; {\em A direct Test for Dust-driven Wind Physics}. 
The program investigates the detailed acceleration 
mechanisms of wind outflow in the outer atmosphere, the upper chromosphere 
and inner circumstellar dust envelope (CDE), of the nearby red supergiant. Using the 
exceptional capabilities of the STIS we observe the near-UV spectrum with 
$\lambda$/$\Delta$$\lambda$$\simeq$33,000 between 2275 \AA\, and 3180 \AA\,
with spatially resolved raster scans across the chromosphere at 0, 200, 400, 600, \&
1000 milli-arcseconds (mas) (Visit 1), at 0 \& 2000 mas (Visit 2), 
and at 0 \& 3000 mas (Visit 3). The spectra are observed with a small
aperture of 63 by 200 mas, using exposure times ranging from 500 s at 
200 mas to 8700 s at 3000 mas, yielding good S/N$\geq$20. The spectra 
have been calibrated with CALSTIS v2.12 using the most recently updated 
calibration reference files. Wavelength calibration accuracies are typically 
better than $\sim$1 detector pixel, or 1.3 $\rm km\,s^{-1}$.

\begin{figure}[t]
  \begin{center}
    \epsfig{file=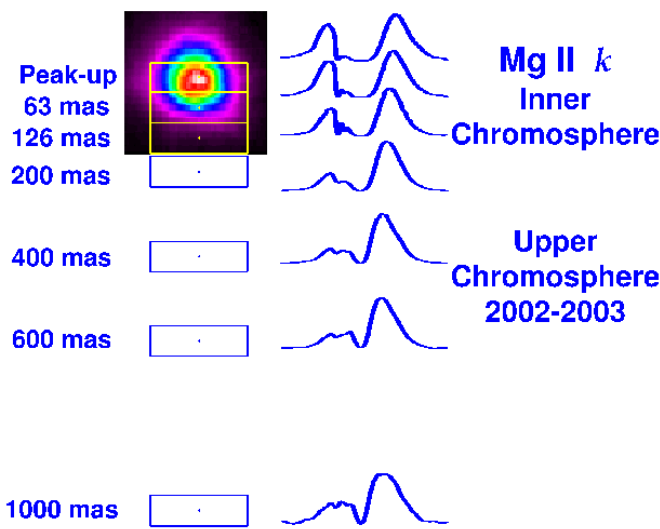, width=9cm}
  \end{center}
%\resizebox{\hsize}{!}{\includegraphics{gravityRGBs.ps}}
\caption{Spatially resolved STIS observations of the Mg~{II} $k$ 
resonance emission line across $\alpha$ Ori's inner and upper chromosphere
using the 63 by 200 mas aperture with respect to a false color near-UV image 
of the inner chromosphere. The Mg~{II} line is scaled to the same
intensity to show important profile shape changes.
\label{fig1}}
\end{figure}

In previous work we modeled the detailed shape of the Mg {\sc ii} $h$ \& $k$
resonance emission lines (Lobel \& Dupree 2000). The lines have previously 
been observed (April 1998) by scanning across the inner chromosphere at 
0, 63, and 126 mas, also using a slit size of 63 $\times$ 200 mas ({\em Table 1}).
Figure 1 shows the Mg {\sc ii} $k$ line profile at the
respective slit positions up to 1000 mas, compared to an image of the near-UV continuum
(in false colors) observed with HST-$FOC$ (Lobel 2003a). The FWHM of the image is about twice
the optical diameter of the supergiant of 56 mas. The central (self-) absorption core
results from scattering opacity in the chromosphere. The asymmetry
of the emission line component intensities probes the chromospheric flow dynamics in our 
line of sight. The profiles observed in the upper chromosphere reveal an increase
of asymmetry with a weaker short-wavelength emission component. It signals substantial 
wind outflow opacity in the upper chromosphere, which fastly accelerates beyond
200 mas ($\simeq$8.1 $\rm R_{*}$). Figure 2  shows an image of Betelgeuse's
CDE observed with MMT-$MIRAC2$ in infrared light around 10 $\mu$m. The outer intensity
contour is drawn at $\sim$1\% of the non-interfered stellar peak intensity ($\sim$4000
mas) from the star's position ({\em at cross}) in this interferometrically nulled 
image (Hinz et al. 1998). The relative target positions (TP) of the $STIS$ 
aperture across the inner CDE are indicated. 

\begin{figure}[t]
  \begin{center}
    \epsfig{file=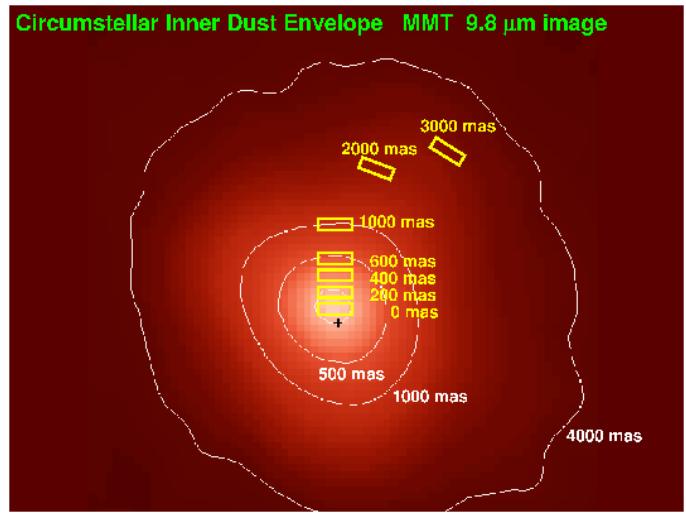, width=9cm}
  \end{center}
%\resizebox{\hsize}{!}{\includegraphics{gravityRGBs.ps}}
\caption{Relative aperture positions of spatially resolved
  spectroscopic observations with STIS in 2002-2003 with respect to a 9.8 $\mu$m
 image of $\alpha$ Ori's inner circumstellar dust envelope from MMT-MIRAC2
  (image adapted from Hinz et al. 1998). \label{fig2}}
\end{figure}
 
\begin{figure*}[t]
  \begin{center}
    \epsfig{file=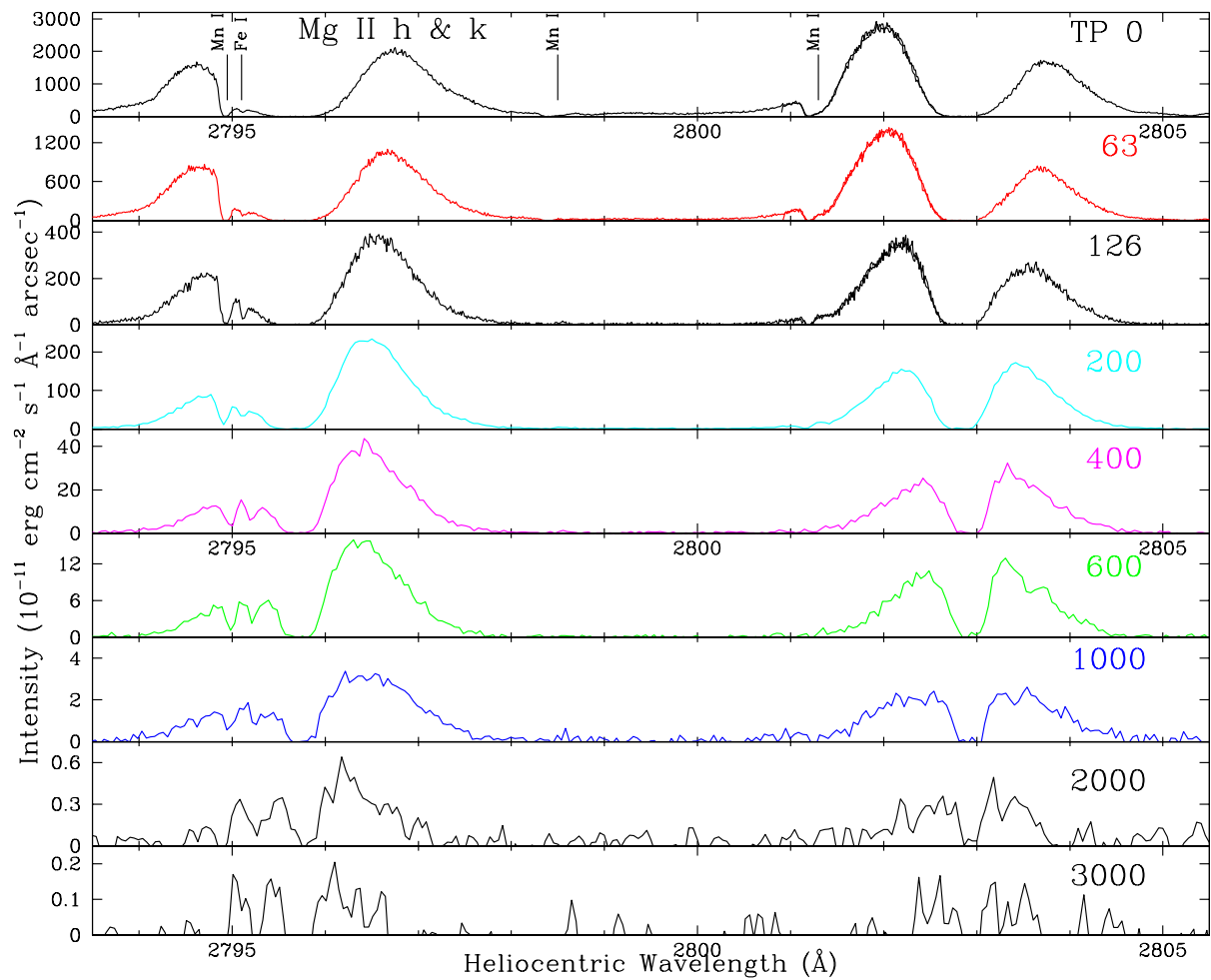, width=13cm}
  \end{center}
%\resizebox{\hsize}{!}{\includegraphics{gravityRGBs.ps}}
\caption{Spatially resolved STIS spectra of Mg~{II} $h$ \& $k$ emission
lines from HST Cycles 7 \& 11, observed off-limb in the chromosphere 
and circumstellar dust envelope of Betelgeuse with the 0.2$\times$0.06 aperture 
out to 3 arcseconds. \label{fig3}}
\end{figure*}    

\tabcolsep0.16cm 
\begin{table*}
\begin{center}
\caption{Spatially resolved spectroscopic exposures of $\alpha$ Ori observed
  with STIS during HST Cycle 7 (GO 7347), using high resolution grating E230H and 
the 0.2$\times$0.06 aperture. Other STIS observations of Jan., Apr., \& Sep. 1998, and March
  1999 using the 0.1$\times$0.03 aperture are not listed. \label{tbl-1}}
\begin{tabular}{ccccccc}
Obs. date   & Dataset     & POS TARG          & Total exp. time & Science exp. & Aperture \\
            & (MAST No.)  & (arcsec)          & (s)             & (amount)&  (name)   \\ 
Apr. 1 1998 &  O4DE05090  &     $-$0.126, 0.0 &   460   &            1    &       0.2$\times$0.06 \\
            &  O4DE050A0  &     $-$0.063, 0.0 &   360   &            1    &       0.2$\times$0.06 \\
            &  O4DE050B0  &        0.000, 0.0 &   260   &            1    &       0.2$\times$0.06 \\
            &  O4DE050C0  &        0.063, 0.0 &   360   &            1    &       0.2$\times$0.06 \\
            &  O4DE050D0  &        0.126, 0.0 &   450   &            1    &       0.2$\times$0.06 \\
\end{tabular}
\end{center}
\end{table*}

\section{Wind Acceleration in the Upper Chromosphere}
\label{sec:win}

Figure 3 shows the detailed line profiles of the broad Mg~{\sc ii} $h$ \&
$k$ emission lines formed in the inner and upper chromosphere 
of Betelgeuse. The resonance emission lines are observed up to 3\arcsec\,
away from the supergiant photosphere, corresponding to $\sim$120 $\rm R_{*}$. 
The upper panel of Fig. 4 compares the Mg {\sc ii} $k$ line profiles 
across the inner and upper chromosphere in velocity scale. The line fluxes 
are scaled to the maximum flux of the long-wavelength emission component. 
We observe a strong decrease of the short-wavelength emission component 
compared to the long-wavelength component by scanning outwards (toward larger TPs), 
which we also observe in resonance lines of Si~{\sc i} $\lambda$2516 
and $\lambda$2507, and in Mg~{\sc i} $\lambda$2852. The short-wavelength 
emission component of the Mg~{\sc ii} $k$ line is however blended with a 
weak Fe~{\sc i} line and a chromospheric Mn~{\sc i} line that can influence 
the $k$ line asymmetry observed across the inner chromosphere. 
The lower panel of Fig. 4 shows four spatially resolved (scaled) Mg~{\sc ii} $k$
profiles from balloon observations (RASOLBA; Staath \& Lemaire 1995) 
out to 9\arcsec\, above the limb of the average quiet \object{Sun}. 
Similar to Betelgeuse, the FWHM of the Mg~{\sc ii} $k$ (and $h$) line,
and the width of the central self-absorption core, decreases
with larger distances above the solar limb. These changes
result from the decrease of electron density and microturbulence velocity
higher in the upper chromosphere, which decreases the scattering of photons 
along the line of sight. The central absorption cores of the Mg~{\sc ii} lines
in Betelgeuse ($d$$\sim$132 pc) can also have narrow contributions from the
Local Interstellar Medium (LISM), but which remain smaller 
(FWHM$\leq$25 $\rm km\,s^{-1}$) than the width of the central absorption 
cores we observe at 2000 and 3000 mas (Lobel 2003b). 
   
\tabcolsep0.16cm 
\begin{table*}
\begin{center}
\caption{Spatially resolved spectroscopic exposures of $\alpha$ Ori observed
with STIS during HST Cycle 11 (GO 9369), using medium resolution grating E230M with the 0.2$\times$0.06
and 0.1$\times$0.03 apertures. \label{tbl-2}}
\begin{tabular}{ccccccc}
Obs. date    & Dataset    & POS TARG       & Total exp. time & Science exp. & Aperture \\
             & (MAST No.)   & (arcsec)          & (s)             & (amount)
             &  (name)  \\ 
Oct. 14 2002 & O6LX01020  &    0.000, 0.0  &  283   &            1    &     0.1$\times$0.03 \\ 
             & O6LX01030  &    0.200, 0.0  &  600   &            1    &     0.2$\times$0.06 \\
             & O6LX01040  &    0.400, 0.0  & 1000   &            1    &     0.2$\times$0.06 \\
             & O6LX01050  &    0.600, 0.0  & 1849   &            1    &     0.2$\times$0.06 \\
             & O6LX01060  &    1.000, 0.0  & 1400   &            1    &     0.2$\times$0.06 \\
             & O6LX01070  &    1.000, 0.0  & 5800   &            2    &
             0.2$\times$0.06 \\
Mar. 02 2003 & O6LX02060  &    0.000, 0.0  &  300   &            1    &     0.1$\times$0.03 \\
             & O6LX02070  &    2.000, 0.0  & 8663   &            3    &
             0.2$\times$0.06 \\
Apr. 23 2003 & O6LX03060  &    0.000, 0.0  &  300   &            1    &     0.1$\times$0.03 \\
             & O6LX03070  &    3.000, 0.0  & 8663   &            3    &     0.2$\times$0.06 \\
\end{tabular}
\end{center}
\end{table*} 

\begin{figure}[t]
  \begin{center}
    \epsfig{file=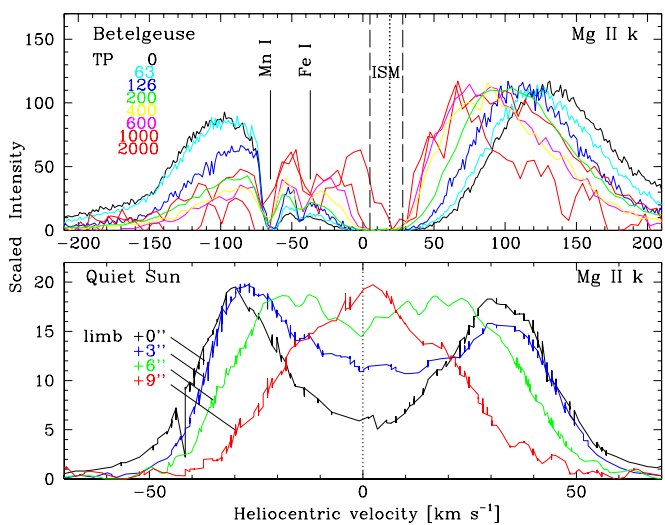, width=7cm}
  \end{center}
%\resizebox{\hsize}{!}{\includegraphics{gravityRGBs.ps}}
\caption{Spatially resolved STIS spectra of the Mg~{II} $k$ emission
line observed across the inner and upper chromosphere of Betelgeuse (upper panel), 
compared to spatially resolved spectra of the $k$ line from balloon
observations (lower panel) above the limb of the average quiet Sun (see text). \label{fig4}}
\end{figure}

\section{Si {\sc i} $\lambda$2516 line profile changes}
\label{sec:sil}
In previous work we modeled the detailed shape of the unblended Si {\sc i} $\lambda$2516 
resonance emission line (Lobel \& Dupree 2001). The line was 
observed by scanning across the inner chromosphere at 0, 25, 50, and 75 mas, 
using an aperture size of 30 $\times$ 100 mas. Figure 5 shows double-peaked
Si~{\sc i} line profiles observed across the inner chromosphere in March
1999. The central (self-) absorption core results from scattering opacity in 
the inner chromosphere. The asymmetry
of the emission component intensities probes the chromospheric flow dynamics in our 
line of sight. The line profiles of the outer chromosphere beyond 75 mas are observed
with a slitsize of 63 $\times$ 200 mas ({\em Table 2}). The profiles appear 
red-shifted and reveal weaker short-wavelength emission components by scanning
outwards. It signals substantial wind outflow opacity in the upper
chromosphere, which fastly accelerates beyond 200 mas. 
The shape of these unsaturated Si~{\sc i}
emission lines is very opacity sensitive to the local chromospheric velocity
field in the line formation region. Similar as for the Mg~{\sc ii} lines, 
the outward decrease of intensity of the 
short-wavelength emission component signals fast acceleration of wind outflow in the 
upper chromosphere. Previous detailed radiative transfer modeling of
Si~{\sc i} $\lambda$2516 revealed that $\alpha$ Ori's inner chromosphere oscillates
non-radially in September 1998 ({\em see Sect. 6}).

\begin{figure}[t]
  \begin{center}
    \epsfig{file=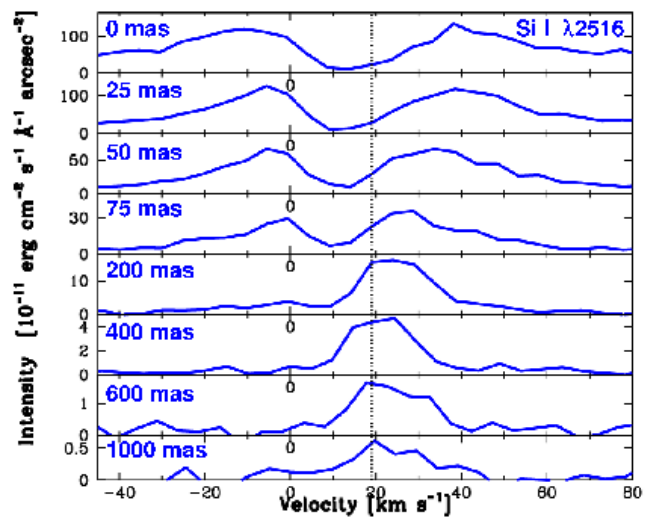, width=9cm}
  \end{center}
%\resizebox{\hsize}{!}{\includegraphics{gravityRGBs.ps}}
\caption{The detailed shape of the unblended Si~{I} $\lambda$2516
emission line reveals a strong increase of profile asymmetry by scanning from the inner 
to the outer extended chromosphere of $\alpha$ Ori. The violet emission wing of the line strongly 
decreases beyond 200 mas due to increased blueshifted scattering opacity in an 
expanding upper chromospheric wind. \label{fig5}}
\end{figure}

\begin{figure}[t]
  \begin{center}
    \epsfig{file=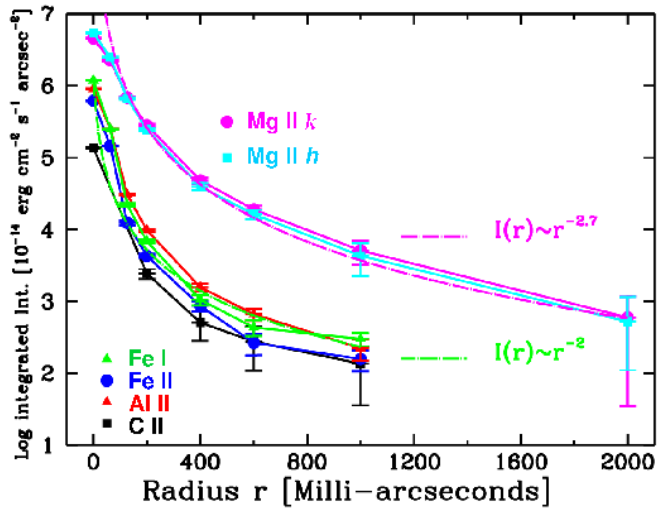, width=9cm}
  \end{center}
%\resizebox{\hsize}{!}{\includegraphics{gravityRGBs.ps}}
\caption{ Integrated intensities of the Mg~{II} resonance emission lines
observed in spatially resolved STIS spectra of Betelgeuse's upper atmosphere. 
A best fit to the radial intensity distributions is obtained for
$I(r)$$\sim$$r^{-2.7}$, signaling important radiative transfer effects on the detailed line
profile formation (see Fig. 3). The flux in the Mg~{\sc ii} emission lines
decreases by a factor of $10^{4}$ compared to the flux at chromospheric disk center.
\label{fig6}}
\end{figure}

\section{Radial Intensity of Chromospheric Lines}
\label{sec:lin}

We observe various ion emission lines as Fe~{\sc ii} $\lambda$2716 (UV 62), Al~{\sc ii} ]
$\lambda$2669 (UV 1), and C~{\sc ii} $\lambda$2327 (UV 1) out to 1\arcsec\, in the 
upper chromosphere (see Fig. 3 of Lobel et al. 2003c).
We select unblended single-peaked emission lines without a central
self-absorption core to determine the radial intensity distribution $I$(r) across the
chromosphere from wavelength integration beyond the line wings. 
Figure 6 compares the $I$(r) of the three ion lines and of Fe~{\sc i} $\lambda$2823
(UV 44) with the intensity distribution observed for the Mg~{\sc ii} $h$ \&
$k$ lines. The lines of the inner chromosphere were observed in April 1998 with 
R$\sim$114,000 at 0, 63, and 126 mas, while the lines in the upper
chromosphere (r$\geq$200 mas) have been observed with medium resolution 
in fall 2002. The spatial scans are however observed with the same 
slitsize of 63 $\times$ 200 mas, so that integrated line intensities 
can be compared. The intensity errorbars are computed from the $STIS$ pipeline 
flux calibration errors, while the radius errorbars are derived from the
projected slitwidth. We observe that the $I$(r) of single-peaked
neutral and ion emission lines are very similar across the chromosphere. 
The lines become rather optically thin in the upper chromosphere with 
a density dependent $I$(r) that is best fit with $I$(r)$\simeq$const~$\times$~$\rm
r^{-2}$. In general the single-peaked neutral emission lines can be observed 
somewhat farther into the upper chromosphere with larger S/N compared to the 
ion lines, although their $I$(r) do not differ significantly within the errors. 
The $I$(r) distribution for the optically thick and self-absorbed Mg~{\sc ii} 
lines differs however significantly with $I$(r)$\simeq$const $\times$ $\rm r^{-2.7}$.
The steeper intensity distribution (note the logarithmic scale for $I$) signals 
important radiative transfer effects for the formation of the detailed 
Mg~{\sc ii} line profiles.

\begin{figure}[t]
  \begin{center}
    \epsfig{file=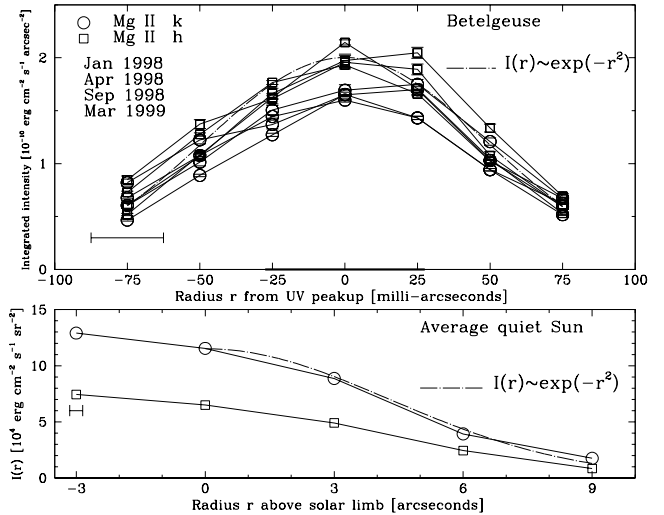, width=7cm}
  \end{center}
%\resizebox{\hsize}{!}{\includegraphics{gravityRGBs.ps}}
\caption{  A comparison of integrated line intensities of Mg~{II}
$h$ \& $k$ derived from spatially resolved observations across the inner 
chromosphere of Betelgeuse ({upper panel}) and the limb of the Sun ({lower panel}).
The off-limb radial intensity distributions follow an exponential dependence,
despite the large difference in chromospheric extension for Betelgeuse and
the Sun.\label{fig7}}
\end{figure}

The upper panel of Fig. 7 shows the $I$(r)-distribution for the Mg~{\sc ii} lines
in four spatially resolved $STIS$ observations during 1998-99. The raster scans 
have been observed with the smallest aperture of 30 $\times$ 100 mas across 
the inner chromosphere of Betelgeuse out to r=$\pm$75 mas. The average
$I$(r) is best fits with an exponential distribution of $I$(r)$\sim$exp($-$$\rm r^{2}$)
({\em dash-dotted line}). In the lower panel of Fig. 7 we wavelength integrate the Mg~{\sc ii} 
line intensities observed above the limb of the average quiet Sun ({\em 
lower panel of Fig. 4}). We also obtain a best fit for an exponential
radial intensity dependence of these optically thick resonance emission lines
above the solar limb. Since it has been suggested that the solar magnesium chromosphere 
can be sustained by the dissipation of mechanical waves, the $STIS$ observations
of Betelgeuse can provide new indications that similar heating mechanisms        
are viable for heating its inner chromosphere. Conversely, the tremendously
large difference observed for the chromospheric extension of the Sun (0.01 to 0.02
$\times$ $\rm R_{\odot}$) compared to Betelgeuse ($\sim$120 $\times$ $\rm
R_{*}$), signals that the efficiency of these chromospheric heating mechanisms 
is very different and strongly determined by the outward acceleration
mechanism(s) in the (magnetic) upper photosphere of a yellow dwarf star 
compared to the very extended (presumably non-magnetic) atmosphere of a red supergiant star. 

\begin{figure}[t]
  \begin{center}
    \epsfig{file=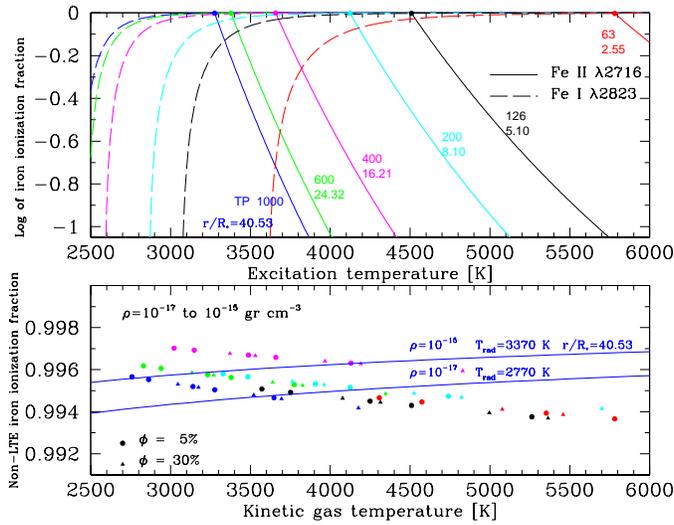, width=7cm}
  \end{center}
%\resizebox{\hsize}{!}{\includegraphics{gravityRGBs.ps}}
\caption{ Iron ionization fractions computed from Fe~{I} and Fe~{II} emission 
line intensities observed out to 1\arcsec\, in the chromosphere of $\alpha$ Ori
(dots in upper panel). The corresponding non-LTE ionization fractions are
computed (lower panel) for 5 gas density values (filled dots and triangles)
with a diluted radiation field of 3000 K, yielding gas kinetic temperatures 
above 2600 K. Colors in upper and lower panels correspond (see text).  
\label{fig8}}
\end{figure}  

\section{Semi-Empiric Model for The Chromosphere}
\label{sec:rnl}

In the upper panel of Fig. 8 we compute the iron ionization fraction
from the radial intensity distribution of the single-peaked
Fe~{\sc i} $\lambda$2823 ({\em dashed lines})
and Fe~{\sc ii} $\lambda$2716 ({\em solid line}) emission lines. 
The intersection of the ionisation curves ({\em solid dots}) provides the 
excitation temperature corresponding to the observed line intensity ratios 
for spontaneous emission in these (optically thin) line transitions for TPs 
up to 1000 mas. We assume the iron solar abundance in our computations. 
The lower panel in Fig. 8 shows these iron non-LTE ionization fractions between 99.3\% and 
99.7\%, which correspond to kinetic gas temperatures between 2600 K and 5800 K,
we compute for five local gas densities $\rho$ for warm chromospheric plasma 
between $10^{-17}$ and $10^{-15}$ $\rm gr\,cm^{-3}$ ({\em filled colored 
symbols correspond to line colors in the upper panel}). 
This gas temperature range corresponds to partial non-LTE iron ionization 
due to a diluted radiation field of $T_{\rm rad}$$\simeq$3000 K ({\em 
e.g. full drawn lines are computed for TP 1000 where $\rm r/R_{*}$=40.53}), 
typical for the upper chromosphere of Betelgeuse.  
The graphs are computed with volume filling factors $\phi$ for warm
chromospheric plasma of 5\% ({\em dots}) and 30\% ({\em triangles}).
We compute that hydrogen is almost neutral for these conditions in 
the upper chromosphere. 

We compute that the warm chromospheric plasma
assumes kinetic gas temperatures above 2600 K out to 1\arcsec\, for 
a range of realistic gas density values we determine in the next Section 
from detailed radiative transport models of the inner CDE. These
chromospheric gas densities correspond to the range of 
model electron densities $N_{e}$ required to compute the self-absorbed 
shape of the Si~{\sc i} $\lambda$2516 resonance line in Fig. 9.
The best fit ({\em solid red line in upper panel})   
to the spatially resolved line profile observed at 75 mas ({\em solid blue
 line}) is obtained for a thermal model of the inner chromosphere (r$\leq$ 10
$\rm R_{*}$) ({\em lower left panel}) with $T_{\rm gas}$ $\leq$5500 K (where 
$N_{e}$$\simeq$2.8 $10^{8}$ $\rm cm^{-3}$) and $N_{e}$ decreases outward. 
The dynamic model ({\em lower panel right}) requires projected
microturbulence velocities with a local maximum of $\sim$13 $\rm km\,s^{-1}$, 
and an outward accelerating (gas) wind velocity increasing to 4.5 $\rm km\,s^{-1}$,  
to match the asymmetric shape and width of the Si~{\sc i} emission line.   
A more extensive discussion on the determination of the semi-empiric model for the inner
chromosphere of Betelgeuse with detailed radiative transfer fits 
is provided in Lobel \& Dupree (2001). Comprehensive animations demonstrating 
important radiative transfer effects on the formation of the detailed Si~{\sc i} line 
profile by varying the log($N_{e}$), $V_{\rm turb}$, and $V_{\rm wind}$
chromospheric model structures are available with this paper, at 
{\tt cfa-www.harvard.edu/$\sim$alobel}, and in the Cool Stars 13 Conference
website at {\tt www.hs.uni-hamburg.de/cs13/ \\
day5/04$\_$Lobel.ppt}

\begin{figure}[t]
  \begin{center}
    \epsfig{file=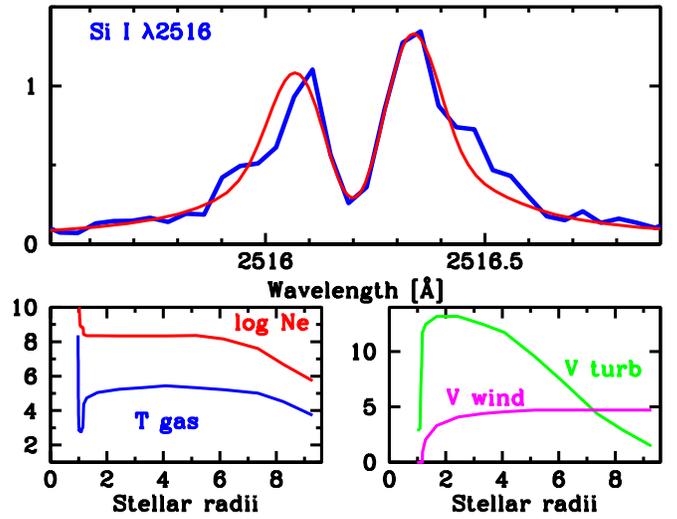, width=6.9cm}
  \end{center}
%\resizebox{\hsize}{!}{\includegraphics{gravityRGBs.ps}}
\caption{Detailed non-LTE radiative transfer fits (red line in upper panel) to
  the spatially resolved Si~{I} $\lambda$2516 emission line observed at 75
  mas (blue line). The corresponding thermal and dynamic structures of the
  inner chromosphere model (lower panels) are varied in movie sequences to
  illustrate the goodness of fit. $T_{gas}$ is shown in kK units, and Vwind and Vturb in 
  km $s^{-1}$ units. \label{fig9}}
\end{figure}

\section{Semi-Empiric Model of Inner Dust Envelope}
\label{sec:rnl}

We determine the model parameters of Betelgeuse's inner CDE 
with radiative transfer in spherical geometry using the {\sc Dusty} 
code (Ivezi\'{c} \& Elitzur 1995; Lobel et al. 1999). 
Figure 10 shows the best fit to the weak silicate dust emission feature 
observed at 9.8 $\mu$m with $IRAS$ (or $ISO$) ({\em red solid lines}). 
The detailed (Kurucz) photospheric input model with $T_{\rm eff}$=3500 K 
and log($g$)=$-$0.5 is processed through a model of the CDE (Lobel et
al. 2000). The best fit yields an inner dust condensation radius 
$R_{c}$$\simeq$573 mas ($\sim$23.2 $\rm R_{*}$), where $\rho_{\rm gas}$$\sim$5 
$10^{-16}$ $\rm gr\,cm^{-3}$ for the cool ambient gas component.
We assume the canonical value of 200 for the gas-to-dust density ratio, although 
an order of magnitude larger can still be adopted for this cool supergiant star. 
The best model fit yields a silicate dust condensation temperature of
$T_{\rm dust}$$\leq$600 K ({\em blue solid dotted lines}), composed of olivine grains
with a size distribution $n(a)$$\sim$$a^{-3.5}$, and 0.05 $\mu$m $\leq$$a$$\leq$ 0.5 $\mu$m. 
We compute that the 9.8 $\mu$m flux optical depth of $\alpha$ Ori's inner CDE is small 
with $\tau_{9.8}$=0.015. Figure 11 shows the corresponding best model fit
({\em green line}) to the dust surface brightness of the inner CDE observed in the 10
$\mu$m MMT images out to 4\arcsec. The relative dust emission intensities 
computed with the model in front of the inner dust cavity ({\em vertical dashed line})  
are not shown. Danchi at al. (1994) find from visibility curves 
very little dust radiation coming from distances below 950 mas, while 
Sutton et al. (1977) observe that less than 20 \% of the 
excess radiation at 10 $\mu$m is emitted by dust between 6~$\rm R_{*}$ and
12~$\rm R_{*}$ from $\alpha$ Ori.  

\begin{figure}[t]
  \begin{center}
    \epsfig{file=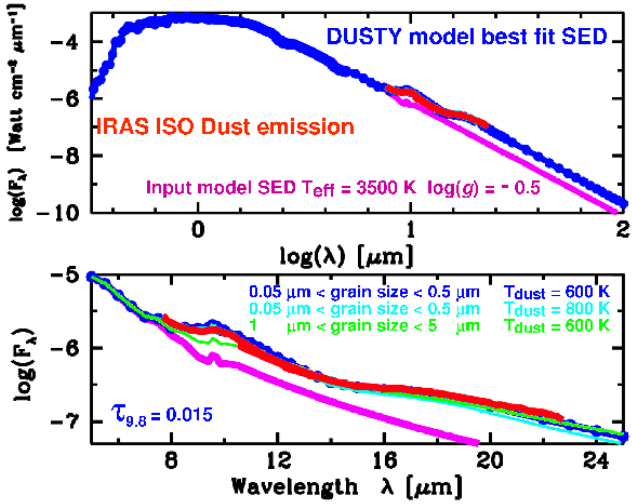, width=9cm}
  \end{center}
%\resizebox{\hsize}{!}{\includegraphics{gravityRGBs.ps}}
\caption{Best fit with {\sc Dusty} (dotted line in upper panel) to the 
IRAS 9.8~$\mu$m silicate dust emission of $\alpha$ Ori (red line). The detailed 
stellar input spectrum (magenta line) is processed through the model of the
circumstellar dust envelope (see Fig. 12). The best fit is obtained for 
olivine grain composition (blue line in lower panel) (see text).
\label{fig10}}
\end{figure}

The upper panel of Fig. 12 shows the kinetic gas temperature structure of the 
warm chromospheric plasma computed with $\rho_{\rm gas}$=5 $\rm 10^{-16}$ 
$\rm gr\,cm^{-1}$ for $\phi$= 1\% and 100\% ({\em bold red lines}). 
The model for the inner chromosphere is computed with detailed
radiative transfer fits to H$\alpha$ and Mg~{\sc ii} (Lobel \& Dupree 2000). 
We compute that the temperatures of the upper chromospheric plasma do not 
decrease to below 2600 K in Sect. 6. The lower panel of Fig. 12 shows the 
model of the CDE we compute from the best fit to the silicate dust emission
observed in the upper chromosphere. To condense dust grains out of
the gas phase the ambient gas temperature remains below the dust 
temperature $T_{\rm dust}$$\leq$600 K ({\em magenta line}). Hence warm
chromospheric plasma must co-exist with cool gas of $T_{\rm gas}$$\leq$600 K beyond 600
mas (Lobel et al. 2004). Radiation pressure onto dust accelerates the grains to a terminal (dust)
outflow velocity of $\sim$13~$\rm km\,s^{-1}$ ({\em green line}). The dust-gas
interaction causes the wind acceleration by dust-gas drag forces. The
acceleration of the warm wind is observed with HST-$STIS$ in the upper
chromosphere of Betelgeuse from the increase of emission line asymmetry 
({\em Sect. 4}). 
 
\begin{figure}[t]
  \begin{center}
    \epsfig{file=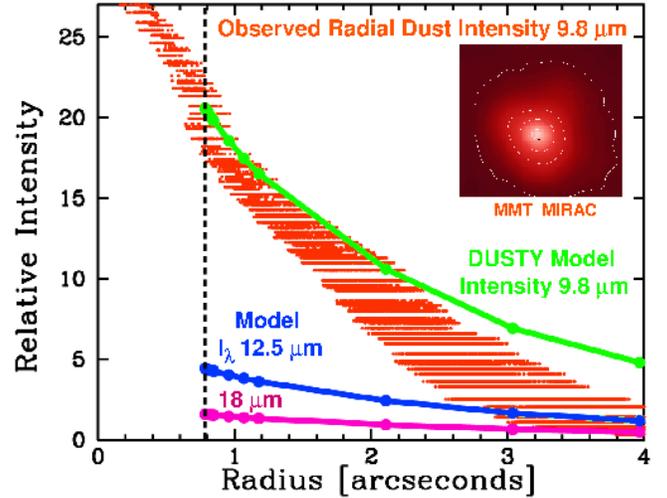, width=9cm}
  \end{center}
%\resizebox{\hsize}{!}{\includegraphics{gravityRGBs.ps}}
\caption{Best fit with {\sc Dusty} to the observed 9.8 $\mu$m radial intensity
 distribution of $\alpha$ Ori's inner dust envelope from MMT-MIRAC2
 images. Model $I_{\lambda}$'s at 12.5 and 18 $\mu$m are also shown.\label{fig11}}
\end{figure}

\begin{figure*}[t]
  \begin{center}
    \epsfig{file=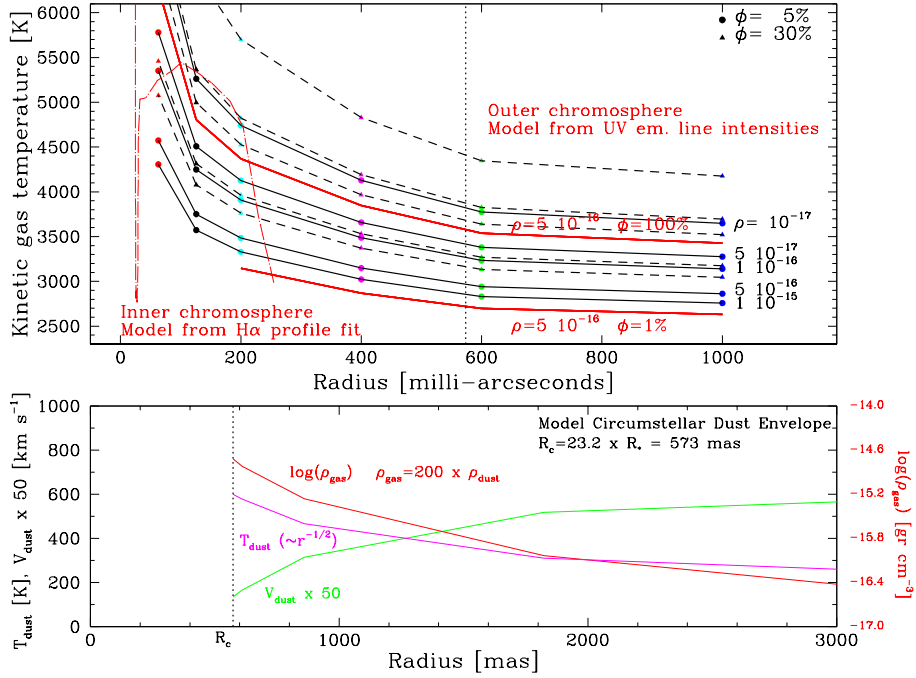, width=9cm}
  \end{center}
%\resizebox{\hsize}{!}{\includegraphics{gravityRGBs.ps}}
\caption{Kinetic gas temperatures computed for the upper chromosphere of
  Betelgeuse with 5 gas densitiy values $10^{-17}$ $\leq$ $\rho$ $\leq$ $10^{-15}$ gr $cm^{-1}$ and volume 
filling factors for chromospheric plasma $\phi$ of 5\% (dots in upper panel) and 30\%
  (triangles). Upper chromospheric Tgas-values with $\phi$=1\% and 100\% 
for $\rho$ = 5 $10^{-16}$ gr $cm^{-1}$ (bold red lines) remain above 2600 K. 
Best fits to dust emission in the upper chromosphere require model 
temperatures (magenta line in lower panel) of ambient gas below 600 K to 
sustain the presence of silicate dust (see text). \label{fig12}}
\end{figure*}

\section{Conclusions}

\begin{itemize}

\item
Based on the Mg~{\sc ii} $h$ \& $k$ emission lines we provide 
the first evidence for the presence of warm chromospheric plasma up to 
three arcseconds away from the star at $\sim$120 $\rm R_{*}$ 
(1 $\rm R_{*}$$\simeq$700 $\rm R_{\odot}$). Other strong emission lines 
of Fe~{\sc ii} $\lambda$2716, C~{\sc ii} $\lambda$2327, and Al~{\sc ii} ] 
$\lambda$2669 are significantly detected out to a full arcsecond. The $STIS$ spectra 
reveal that Betelgeuse's upper chromosphere extends far beyond the circumstellar 
H$\alpha$ envelope of $\sim$5 $\rm R_{*}$, determined from previous
ground-based imaging (Hebden et al. 1987). 

\item
The flux in the broad and self-absorbed resonance lines of Mg~{\sc ii} 
decreases by a factor of $10^{4}$ compared to the flux at chromospheric disk 
center. We observe strong asymmetry changes in the Mg~{\sc ii} $h$ \& $k$,  
Si~{\sc i} $\lambda$2516, and $\lambda$2507 resonance emission line
profiles when scanning off-limb, signaling outward acceleration of gas 
outflow in the upper chromosphere. It directly demonstrates that
the warm chromospheric wind of Betelgeuse extends far beyond the dust
condensation radius determined from stationary dust-driven wind
acceleration models. 

\item
A detailed comparison of profile changes observed for the Mg~{\sc ii} $h$ \& $k$
lines with HST-$STIS$ across the chromosphere of Betelgeuse reveals a striking resemblance 
to the emission line profile changes observed across the limb of the average
quiet Sun. We observe that the integrated radial intensities $I$(r) of the
Mg~{\sc ii} lines follow an exponential distribution 
($I$(r)$\sim$exp($-$$\rm r^{2})$) for the magnesium chromosphere of the Sun
and for the inner chromosphere of Betelgeuse out to a distance r of $\sim$75 mas. 
On the other hand, our measurements reveal that the integrated line intensities  
with distance in the {\em upper} chromosphere beyond 75 mas cannot be modeled
with an exponential distribution, and require a sharply peaked function of
$I$(r)$\sim$$\rm r^{-2.7}$. 

\item
We compute that the local kinetic gas temperatures of the warm 
chromospheric gas component in the outer atmosphere exceed 2600~K, 
when assuming local gas densities of the cool gas component we 
determine from radiative transfer models that fit the 9.8 $\mu$m 
silicate dust emission and its radial surface brightness. 
The spatially resolved $STIS$ spectra directly demonstrate that 
warm chromospheric plasma co-exists with cool 
gas in Betelgeuse's circumstellar dust envelope. 
The recent STIS data can therefore support the dust-gas interaction
driving mechanism, yielding the high mass-loss rates observed in cool
supergiants and AGB stars. From a thermodynamic point of view however,
the uniform dusty wind model fails to account for the large local
kinetic gas temperature differences because the warm chromospheric gas
($T_{\rm gas}$$\geq$ 2600 K) is observed far inside the CDE ($T_{\rm dust}$ $\leq$ 
600 K), out to 3\arcsec\,, $\sim$120 $\rm R_{*}$ away from Betelgeuse.

\end{itemize}

\begin{acknowledgements}

This reseach is based on data obtained with the NASA/ESA Hubble Space
Telescope, collected at the STScI, operated by AURA Inc., under contract
NAS5-26555. Financial support has been provided by STScI grant HST-GO-09369.01.

\end{acknowledgements}

\end{document}